\documentstyle[twocolumn,epsf,aps]{revtex}
\draft
\begin{document}

\title{Generalized mean-field study of a driven lattice gas}
\author{Gy\"orgy Szab\'o and Attila Szolnoki}
\address{Research Institute for Materials Science, H-1525 Budapest,
POB 49, Hungary}

\address{\em \today}
\address{
\centering{
\medskip \em
\begin{minipage}{15.4cm}
{}~~~Generalized mean-field analysis has been performed to study
the ordering process in a half-filled square lattice-gas model with
repulsive nearest neighbor interaction under the influence of a
uniform electric field. We have determined the configuration
probabilities on 2-, 4-, 5-, and 6-point clusters excluding the
possibility of sublattice ordering. The agreement between the
results of 6-point approximations and Monte Carlo simulations
confirms the absence of phase transition for sufficiently
strong fields.
\end{minipage}
}}
\maketitle

\narrowtext

\section{INTRODUCTION}
\label{sec:intro}

The driven lattice-gas models have been extensively investigated in
order to understand the effect of electric field on the ordering
processes (for a review see\cite{konyv}). These models were introduced
as a generalization of the traditional lattice-gas models taking into
account the effect of a uniform driving (electric or gravitational)
field on the particle jumps \cite{KLS}. More precisely, beside the
nearest neighbor interactions the particle jumps are affected by a
driving field $E$. In the absence of driving field these systems
undergo a well known phase transition. The external field induces a
material transport through the system, modifies the ordering processes
and generates long range correlations. In this area most of the
research activity has been concentrated on the system with attractive
interaction.

In this work our attention is focused to the half-filled,
two-dimensional lattice gas with repulsive nearest neighbor
interaction. Here we follow the notation given in a previous paper
\cite{poly}. The equilibrium system exhibits a phase transition
(at the N\'eel temperature, $T_{N}$) belonging to the Ising
universality class. According to renormalization group \cite{LSZ} and
mean-field \cite{Dickr} analyses the sublattice ordering remains
continuous for weak driving fields while $T_{N}$ decreases with $E$.
The sublattice ordering becomes first order above a threshold field
and vanishes for $E>2$ (in unit defined later). The mean-field
analysis has demonstrated clearly that both the field-dependence of
critical temperature and the value of threshold field are affected
by the choice of jump rate. For example, the threshold field is about
1 or 1.5 when using Metropolis or Kawasaki rates respectively. Some
features of the predicted phase transitions were checked by Monte
Carlo (MC) simulations performed on lattices whose sizes were smaller
than $60 \times 60$ \cite{konyv,LSZ}. These system sizes, however,
are proved to be too small to exhibit the relevant behavior.

Using MC simulations in a previous work we have shown that instead
of the homogeneous ordered state a self-organizing polydomain state
will be stable at low temperatures for sufficiently large system
sizes \cite{poly}. Several authors have observed a nucleation process
which destroys the monodomain state into a polydomain one
\cite{poly,Teitel}. This means that the monodomain structure is
unstable in the thermodynamic limit.
The above phenomena are explained as a consequence of the enhanced
particle transport along the domain boundaries leading to interfacial
instability. More precisely, a simple phenomenological model was able
to describe both the interfacial instability \cite{poly} and the
existence of a critical nucleon size which depends on field strength.

The translation symmetry is not broken in the polydomain structure at
low temperatures therefore this state is considered as an analytical
continuation of the high temperature one. This fact inspired us to
reinvestigate the earlier mean-field analysis based on the hypothetical
stability of the symmetry-breaking ordered state at low temperatures.
Now we will suppose that the translation invariant state is stable for
arbitrary temperatures.

The dynamical mean-field technique was used by Dickman to evaluate
the average sublattice occupations and the probability of all the possible
configurations on neighboring sites oriented horizontally and vertically
\cite{Dickr}. The time-consuming extension of this method for larger
clusters is straightforward. These extended methods proved to be very
successful for the one-dimensional systems, particularly for a driven
lattice gas \cite{1ddlg} and stochastic cellular automata \cite{CA}.

In this work the above model is studied with determining the
configuration probabilities on 2-, 4-, 5-, and 6-point clusters. Details
of such a calculation are given in previous papers
\cite{Dickr,1ddlg,CA,Dicka,torl} therefore now we concentrate
on the discussion of this series of calculations. Within this approach
the appearance of polydomain structure is indicated by the strengthening
of short range order, i.e., the configurations corresponding to the
ordered particle distributions become dominant on the given cluster.

The output of mean-field calculation is easily comparable with the results
of MC simulations. Systematic MC simulations on sufficiently large
systems were performed only for $E=0.4$. Now we have carried out
simulations for larger fields. The present MC results confirm our
previous conclusions and support the predictions of 6-point
approximations for sufficiently large fields when the typical domains
are not significantly larger than the cluster used in our 6-point
approximation.

\section{THE METHOD}
\label{sec:form}

In the driven lattice-gas model the interacting particles are distributed
on the sites of a square lattice imposing periodic boundary conditions.
The particle distribution is described by a set of occupation variables
$n_i$ taking the values 1 or 0 if a particle is present or absent at site $i$.
In the half-filled ($c=1/2$) system the particles can jump to one of the
empty nearest neighbor sites with a probability depending on the energy
difference between the final and initial positions as formulated in
Ref.\cite{poly}. To avoid the difficulties come from the non-analytic
feature of Metropolis rate we used Kawasaki rate. Following the
traditions the strength of the repulsive nearest neighbor pair interaction
is chosen to be unity. The strength of vertical electric field is measured
in such a unit expressing the potential energy variation for a jump along
the field. For horizontal jumps the driving field does not modify the jump
rates corresponding to detailed balance.

At the level of $k$-point approximation the translation invariant stationary
particle distribution is characterized by the quantity $p_k(n_1,\ldots ,n_k)$
which describes the probability of a given $(n_1,\ldots ,n_k)$ configuration
on a compact cluster of $k$ sites. These quantities satisfy consistency
conditions related to translation invariance \cite{1ddlg,CA} and other
symmetry relations \cite{konyv}. The approach
involves finding a hierarchy of equations of motion for these configuration
probabilities. The set of these equations is solved numerically in the
stationary state.

Using this technique we have determined the configuration probabilities
at different levels. In 2-point (pair) approximation the configuration
probabilities are evaluated on two neighboring sites oriented horizontally
and vertically. Due to the consistency conditions
these quantities are described by introducing only two parameters
in the half-filled system. For 4-point approximation we need 6 parameters
to characterize the probability of each possible configurations on a
$2 \times 2$ cluster. In the five-point cluster the central site is
surrounded by its nearest neighbors. Finally, different 6-point
approximations are performed using rectangular clusters of both
$2\times 3$ and $3 \times 2$ sites. In these latter cases we had to
determine 16 and 18 independent parameters.
In the present work larger clusters are not investigated because the
number of parameters increases rapidly with the cluster size.

To check these results we have carried out MC simulations with varying
the temperatures $T$ for $E=1$, 1.5, 2, 3, and 4. In these simulations the
system size is chosen to be much larger than the typical domain size
observable when displaying the particle distribution. Namely, the
simulations are performed on a rectangular boxes with sizes of
$256 \times 512$ and $180\times 180$ for $E=1$ and stronger fields
respectively. In each cases we have confirmed that the chessboard-like
initial configuration transforms into a high temperature (or polydomain)
state after some thermalization. Unfortunately, the relaxation time
becomes extremely long for low temperatures. In this situation the
simulation of relaxation process may be accelerated by either generating
several point defects (Frenkel pairs) in the initial ordered state or using
the algorithm described by Sadiq\cite{Sadiq}. These simulations
justified undoubtedly our basic assumption, namely, the absence of
sublattice ordering (long range order) in the stationary state.

In principle, using MC simulations all the cluster configuration probabilities
are easily checkable with a desired accuracy. For simplicity, however, our
comparison will be restricted to the values of $p_{\parallel}(1,0)$ and
$p_{\perp}(1,0)$ denoting the probability of $(1,0)$ configurations on
neighboring sites oriented parallel or perpendicular to the vertical field.
These quantities express the short range order and the difference between
the horizontal and vertical pairs reflects the violation of $x-y$ symmetry.
Furthermore, their sum is directly related to the average internal energy.

\section{RESULTS}
\label{sec:results}

In order to illuminate the predictions of the $k$-point approximations we
report the results in order from $k=2$ to 6. In 2-point approximation our
results agree with those found by Dickman at high temperatures
\cite{Dickr}. Figure 1 illustrates the quantities $p_{\parallel}(1,0)$
and $p_{\perp}(1,0)$ as a function of temperature for some typical values
of $E$. Henceforth the subscript will be omitted if it is possible.

\begin{figure}
\centerline{\epsfxsize=8cm
                   \epsfbox{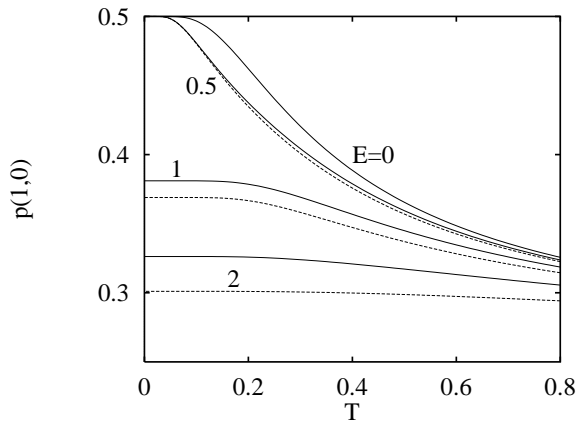}
                   \vspace*{2mm}     }
\caption{Temperature dependence of the probability of $(1,0)$ configurations
on two neighboring sites positioned parallel (solid line) and perpendicular
(dashed line) to the driving field whose strength is indicated by the figures.}
\label{fig:2p}
\end{figure}

In the absence of driving field the parallel and perpendicular directions
are equivalent and $p(1,0)=p(0,1) \to 1/2$ if $T \to 0$. This limit
represents a completely ordered (chessboard) structure.
For low temperatures an estimation may
be given for characterizing the extension of ordered domains along the
principal directions. Namely the probability of finding ordered structure
on $x$ subsequent sites is proportional $\exp (-x/\xi )$ where $\xi =
-1/ \ln [2 p(1,0)]$ defines a characteristic length. In this sense, the
present calculation suggest a polydomain structure with ``typical
domain sizes'' increasing when the temperature is decreased
($\xi \to \infty$ if $T \to 0$).

For low fields the ordering process is slightly modified. The appearance
of large domains are predicted at lower temperature and the ordered regions
are elongated along the field. For the fields $E<1$ the characteristic lengths
($\xi$) diverge along both directions in the limit $T \to 0$. For stronger
fields, however, the formation of large domains is suppressed and the
difference between $p_{\parallel}(1,0)$ and $p_{\perp}(1,0)$ becomes
more striking.

\begin{figure}
\centerline{\epsfxsize=8cm
                   \epsfbox{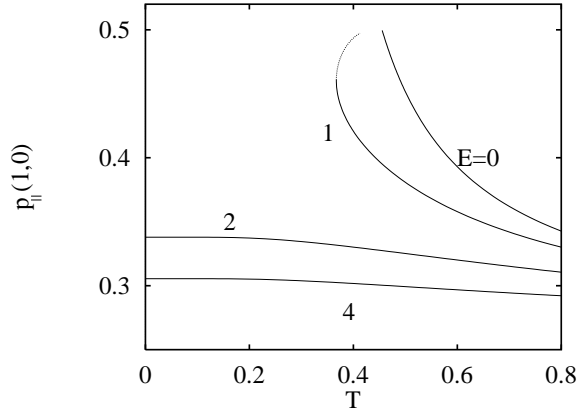}
                   \vspace*{2mm}     }
\caption{Probability of (1,0) configuration along the drive as suggested by
4-point approximations for different fields as indicated. Dotted line
represents unstable solution.}
\label{fig:4p}
\end{figure}

{}From the results of 4-point approximation (see Fig.~\ref{fig:4p}) a
significantly different picture may be concluded. In the high field limit
this approximation confirms qualitatively the previous results. The violation
of $x-y$ symmetry is also recognizable. In spite of our naive expectation
we have not found stable solution satisfying translation invariance at low
temperatures for fields $E<2$. In these cases the numerical solution of the
time-dependent equations indicates that the system develops toward the
chessboard-like distribution. In the close vicinity of the perfectly ordered
state, however, the rounding errors forced the computation to stop. These
observations may be interpreted as arguments to support the sublattice
ordering. In contrary to the early results this approximation predicts an
increase of  N\'eel temperature at low fields. There is an other indication
making the reliability of the 4-point approximation questionable. Namely,
the probabilities of the one- and some two-particle configurations coincide
for $E=0$, while the symmetry relations allow them to be different.

\begin{figure}
\centerline{\epsfxsize=8cm
                   \epsfbox{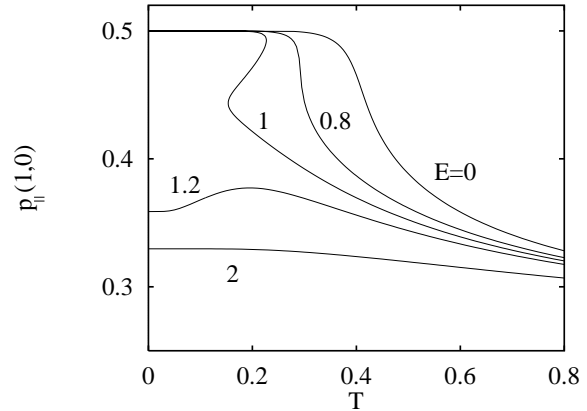}
                   \vspace*{2mm}      }
\caption{Configuration probability $p(1,0)$ vs. temperature for different
fields in 5-point approximation.}
\label{fig:5p}
\end{figure}

To resolve the discrepancy between the suggestions of 2- and 4-point
approximations we have extended the mean-field analysis taking larger
clusters into consideration. The results of 5-point approximation are
plotted in Fig.~\ref{fig:5p}. For low fields the results suggest continuous
transition in qualitative agreement with the prediction of 2-point
approximation. Now the formation of large domains appears at higher
temperatures. This calculation predicts first order transition within a
range of electric field as demonstrated by a typical curve ($E=1$) in
Fig.~\ref{fig:5p}. If $E>1.2$ than the 5-point approximation supports
that the system remains disordered in the zero field limit.

\begin{figure}
\centerline{\epsfxsize=8cm
                   \epsfbox{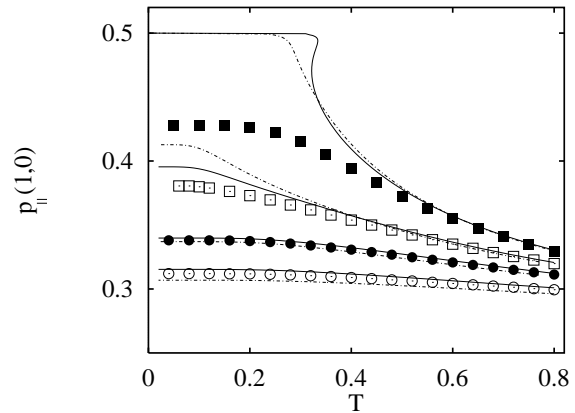}
                   \vspace*{2mm}     }
\caption{Comparison of $p(1,0)$ results of $2 \times 3$- (solid lines) and
$3 \times 2$-point (dashed lines) approximations with MC data
for driving fields $E=1$, 1.5, 2 and 3 (from top to bottom).}
\label{fig:6p}
\end{figure}

Similar set of curves have been obtained when repeating this calculation
at the levels of $2\times 3$- and $3 \times 2$-point approximations. The
quantitative agreement with MC data is satisfactory if $E \ge 1.5$ as
demonstrated in Fig.~\ref{fig:6p}.

Notice that the MC simulations confirm
the absence of phase transitions at the given fields in contrary to early
predictions mentioned in the Introduction. In these cases the correlation
lengths in both directions (defined above) remain finite in the limit
$T \to 0$. For example, decreasing the temperature $\xi_{\parallel}$
goes to 3.6 if $E=1.5$. This value of $\xi$ is close to 3 which is the
longitudinal size of the cluster providing the best agreement.

For weaker fields the MC simulations suggest longer correlation lengths
in the zero temperature limit. For example, $\xi_{\parallel}=6.4$ and
$\xi_{\perp}=5.5$ if $E=1$ and $T=0$. In this case the ``typical domain''
is significantly larger than the clusters used in the present approximations.
These ``typical domains'' (as well as the enhanced material transport along
the interfaces maintaining this structure) appear clearly when visualizing
the particle distribution during the MC simulations. The complicate structure
and pattern formation mechanism can not be taken into account correctly by
such a small clusters. We believe that this is the reason why we have found
substantial differences between the predictions of the generalized
mean-field analysis and the results of MC simulations for $E=1$ and lower
fields in the low temperature region.

Obviously, choosing larger and larger clusters one can increase the accuracy
of the present method and shift the boundary of validity toward lower fields
and temperatures. Unfortunately, the efficiency of this strategy is
questionable because the number of parameters to be determined numerically
increases exponentially with the number of lattice points within the cluster.

Beside the $x$ and $y$ directions this technique distinguishes the forward
and backward directions in the presence of driving field. As discussed in
previous papers \cite{1ddlg,torl} the difference between the forward and
backward directions can not appear explicitly at the level of pair
approximation due to the strong restrictions of consistency conditions. At
higher levels, however, differences can appear between the probabilities
of two distinct configurations considered as reflections with respect to the
transverse symmetry axis of the cluster. In 4-point approximation this type
of symmetry breaking may be described by a parameter which is an odd
function of $E$. At the same time we need 4 parameters to characterize
this symmetry breaking on the $2 \times 3$ cluster. In general we can say
that this forward-backward symmetry breaking is a weak effect. The
corresponding parameters vanish both in the absence of driving field and
in the disordered (high temperature and/or high field) state.

\section{CONCLUSIONS}
\label{sec:conc}

The mean-field analysis of the half-filled two-dimen\-sional driven
lattice-gas model with repulsive interaction has been revised and extended
to 4-, 5-, and 6-point approximations. In the present investigations we
have taken into consideration the results of Monte Carlo simulations.
Varying the temperature at some fixed driving fields these simulations
have justified the absence of sublattice ordering. That means that in the
low temperature stationary state one can observe a self-organizing
polydomain structure. In other words, instead of the long range order
(typical in equilibrium systems) now we can find only short range order.
For such particle distributions the translation invariance of the system is
not broken. As a consequence the mean-field analysis may be simplified
because the low temperature (self-organizing polydomain) structure can
be considered as an analytical continuation of the high temperature state.

The mean-field analysis have been performed at different levels. As
expected the best agreement with MC data is found when using the
largest clusters ($2 \times 3$ and $3 \times 2$) on which the configuration
probabilities are evaluated numerically. For driving fields $E=1.5$, 2,
and 3 the 6-point approximations give adequate description on the effect
of driving field on the ordering process. In these cases the typical area
of ordered regions is small even at zero temperature. More precisely,
it does not exceed the area of cluster we used.

The size difference between the
ordered domains and the $2 \times 3$ cluster becomes more significant
at lower fields because the typical transverse size increases
as $1/E$ \cite{poly}. In this situation the predictions of the 6-point
approximations are no longer valid because it can not take into account
the effect of enhanced interfacial material transport responsible for the
self-organizing domain structure. Since the capability of the
present mean-field analysis is limited therefore we need other approaches
to have a more complete description of this non-equilibrium state.

\acknowledgements

This research was supported by the Hungarian National Research Fund
(OTKA) under Grant Nos. T-4012 and T-16734.

\end{document}